\documentclass[conference,a4paper]{IEEEtran}
\IEEEoverridecommandlockouts

\usepackage{fancyhdr}
\usepackage[dvips]{graphicx}
\usepackage{indent}
\usepackage{amsmath,amssymb}
\usepackage{paralist}
\usepackage{eclbkbox}
\usepackage{subfigure}

\makeatletter

\def\tbcaption{\def\@captype{table}\caption}
\def\figcaption{\def\@captype{figure}\caption}
{\pointlessenum\begin{enumerate}}%
{\end{enumerate}}

\begin{document}

\title{A Generation Method of Filtering Rules of Twitter Via Smartphone Based Participatory Sensing System for Tourist by Interactive GHSOM and C4.5
\thanks{\copyright 2012 IEEE. Personal use of this material is permitted. Permission from IEEE must be obtained for all other uses, in any current or future media, including reprinting/republishing this material for advertising or promotional purposes, creating new collective works, for resale or redistribution to servers or lists, or reuse of any copyrighted component of this work in other works.}
}

\author{\IEEEauthorblockN{Takumi Ichimura}
\IEEEauthorblockA{Faculty of Management and Information Systems,\\
Prefectural University of Hiroshima\\
1-1-71, Ujina-Higashi, Minami-ku,\\
Hiroshima, 734-8559, Japan\\
Email: ichimura@pu-hiroshima.ac.jp}
\and
\IEEEauthorblockN{Shin Kamada}
\IEEEauthorblockA{Graduate School of Comprehensive Scientific Research,\\
Prefectural University of Hiroshima\\
1-1-71, Ujina-Higashi, Minami-ku,\\
Hiroshima, 734-8559, Japan\\
Email: q222003qr@pu-hiroshima.ac.jp}
}

\maketitle
\thispagestyle{plain}

\fancypagestyle{plain}{
\fancyhf{}	% clear all header and footer fields
\fancyfoot[L]{}
\fancyfoot[C]{}
\fancyfoot[R]{}
\renewcommand{\headrulewidth}{0pt}
\renewcommand{\footrulewidth}{0pt}
}

\pagestyle{fancy}{
\fancyhf{}
\fancyfoot[R]{}}
\renewcommand{\headrulewidth}{0pt}
\renewcommand{\footrulewidth}{0pt}

\begin{abstract}
Mobile Phone based Participatory Sensing (MPPS) systems involve a community of users sending personal information and participating in autonomous sensing through their mobile phones. Sensed data can also be obtained from external sensing devices that can communicate wirelessly to the phone. We have developed the tourist subjective data collection system with Android smartphone. The tourist can tweet the information of sightseeing spots by using the application. The application can determine the filtering rules to provide the important information of sightseeing spot. The rules are automatically generated by Interactive Growing Hierarchical SOM and C4.5.
\end{abstract}


\IEEEpeerreviewmaketitle

\section{Introduction}
\label{sec:Introduction}
The current information technology can collect various data sets  because the recent tremendous technical advances in processing power, storage capacity and network connected cloud computing. The sample record in such data set includes not only numerical values but also language, evaluation, and binary data such as pictures. The technical method to discover knowledge in such databases is known to be a field of data mining and developed in various research fields.

Mobile Phone based Participatory Sensing (MPPS) systems involve a community of users sending personal information and participating in autonomous sensing through their mobile phones \cite{Lane2010}. Sensed data can be obtained from sensing devices present on mobiles such as audio, video, and motion sensors, the latter available in high-end mobile phones. Sensed data can also be obtained from external sensing devices that can communicate wirelessly to the phone. Participation of mobile phone users in sensorial data collection both from the individual and from the surrounding environment presents a wide range of opportunities for truly pervasive applications. 

For example, the technology for determining the geographic location of cell phones and other hand-held devices is becoming increasingly available. The tourist subjective data collection system with Android smartphone has been developed \cite{Android_Market}. The application can collect subjective data such as pictures with GPS, geographic location name, the evaluation, and comments in real sightseeing spots where a tourist visits and more than 500 subjective data are stored in the database. Attractive knowledge discovery for sight seeing spots is required to promote the sightseeing industries.

Data mining is the process of applying these methods to data with the intention of uncovering hidden patterns. An unavoidable fact of data mining is that the subsets of data being analyzed may not be representative of the whole domain, and therefore may not contain examples of certain critical relationships and behaviors that exist across other parts of the domain. Moreover, the probability of inclusion of missing data and/or contradictory data becomes high because the means or instrumentality for storing information is to store the raw data in the storage by the automated collecting process. Data mining is seen as an increasingly important tool by modern business to transform unprecedented quantities into business intelligence giving an informational advantage. Some data mining tools are currently used in a wide range of profiling practices such as marketing, fraud detection and medical information. As data sets have growth in size and complexity, automatic data processing technique in the field of computer sciences is required.

In this paper, we examined to classify them to investigate the effectiveness of the interactive GHSOM and the knowledge is extracted by using the classification results of the interactive GHSOM \cite{Ichimura2011a} and C4.5 \cite{Quinlan96}. Rauber et al. proposed the traditional algorithm of the growing hierarchical self organizing map (GHSOM) \cite{Rauber02}. The algorithm has been chosen for its capability to develop a hierarchical structure of clustering and for the intuitive outputs which help the interpretation of the clusters. However, GHSOM divides a data set into sub clusters immoderately if the distribution of samples is complex. There is a trade off for human designers between the investigation of the shape of a partial detailed cluster and the complete control of entire distribution of samples. In order to grasp an overview of tree structure of GHSOM, the interactive GHSOM \cite{Ichimura2011a} has been proposed to restrain the growing of hierarchy in GHSOM by reforming the map in each layer interactively. The criteria for interactive GHSOM are designed by the adjustment of parameter settings according to the squared error and the size of map.

The acquired knowledge via Smartphone based tourist participatory sensing system makes an important role to promote sightseeing spots, because tourist information never seen in web sites or guidebooks can be discovered. In order to inform the travelers the valuable information, our Android application was extended to apply some enhancements on online social networking service and microblogging service. However, a short message to be tweeted is not always appropriate for tourists and the filtering rules to judge that the outgoing message is valuable are required. This paper describes a generation method of filtering rules according to clustering results by the interactive GHSOM and C4.5.

The remainder of this paper is organized as follows. Section \ref{sec:interactiveGHSOM} describes the algorithm of interactive GHSOM and its interface tool. Section \ref{sec:experiments} explains the tourist subjective data. Because the message is too short to extract the important characters, the supplement information should be retrieved from websites. Experimental classification results of MPPS data by interactive GHSOM and C4.5 were described. The filtering rules to find the important message to be tweeted is given. In Section \ref{sec:conclusion}, we give some discussions to conclude this paper.

\section{Interactive GHSOM and Its Android Smartphone Tool}
\label{sec:interactiveGHSOM}

\subsection{Growing Hierarchical SOM}
\label{sec:GHSOM}
\begin{figure}[tb]
\begin{center}
\includegraphics[scale=0.5]{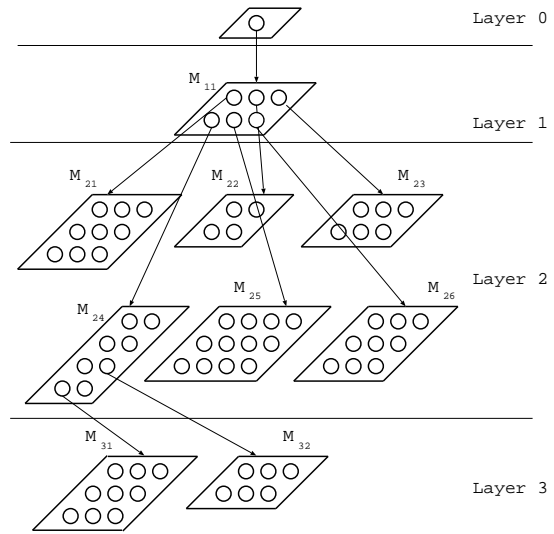}
\caption{A Hierarchy Structure in GHSOM}
\label{fig:overviewGHSOM}
\end{center}
\end{figure}

A basic algorithm of GHSOM was described in \cite{Rauber02}. The algorithm has been chosen for its capability to develop a hierarchical structure of clustering and for the intuitive outputs which help the interpretation of the clusters. These capabilities allow different classification results from rough sketch to very detailed grain of knowledge. This technique is a development of SOM, a popular unsupervised neural network model for the analysis of high dimensional input data \cite{Kohonen95}. Fig.\ref{fig:overviewGHSOM} shows the overview of hierarchy structure in GHSOM.

\subsection{Control of growing hierarchies}
The process of unit insertion and layer stratification in GHSOM works according to the value of 2 kinds of parameters. Because the threshold of their parameters gives a criterion to hierarchies in GHSOM, GHSOM cannot change its structure to data samples adaptively while training maps. Therefore, only a few samples are occurred in a terminal map of hierarchies. In such a case, GHSOM has a complex tree structure and many nodes(maps). As for these classification results, the acquired knowledge from the structure is lesser in scope or effect in the data mining. When we grasp the rough answer from the specification in the data set, the optimal set of parameters must be given to a traditional GSHOM. It is very difficult to find the optimal values through empirical studies.

We propose the reconstruction method of hierarchy of GHSOM even if the deeper GHSOM is performed. A stopping criterion for stratification is defined. Moreover, if the quantization error is large and the condition of hierarchies is not satisfied, the requirements for redistribution of error are defined.

\begin{description}
\item[Case 1)]If Eq.(\ref{eq:ichimuraGHSOM-1}) and the hierarchies are satisfied, stop the process of hierarchies and insert new units in the map again.
\begin{equation}
n_{k}\leq \alpha n_{I},
\label{eq:ichimuraGHSOM-1}
\end{equation}
where, $n_{k},\, n_{I}$ mean the number of input samples for the winner unit $k$ and of the all input samples ${\it I}$, respectively. The $\alpha$ is a constant.
\item[Case2)]If the quantization error is not larger and the addition of layer is not executed, we may meet the situation that the quantization error of a unit is larger than the quantization error in an overall map. If Eq.(\ref{eq:ichimuraGHSOM-2}) is satisfied, then a new unit is inserted.
\begin{equation}
qe_{k}\geq \beta \tau_{1} \sum qe_{y},\, y \in {\bf S}_{k},
\label{eq:ichimuraGHSOM-2}
\end{equation}
where ${\bf S}_{k}$ is the set of winner units $k$. $\beta\tau_{1}$ is a constant for the quantization error.
\end{description}

\subsection{An interface of interactive GHSOM}
\begin{figure}[tbp]
\begin{center}
\subfigure[Simulation Result]{
\includegraphics[scale=0.4]{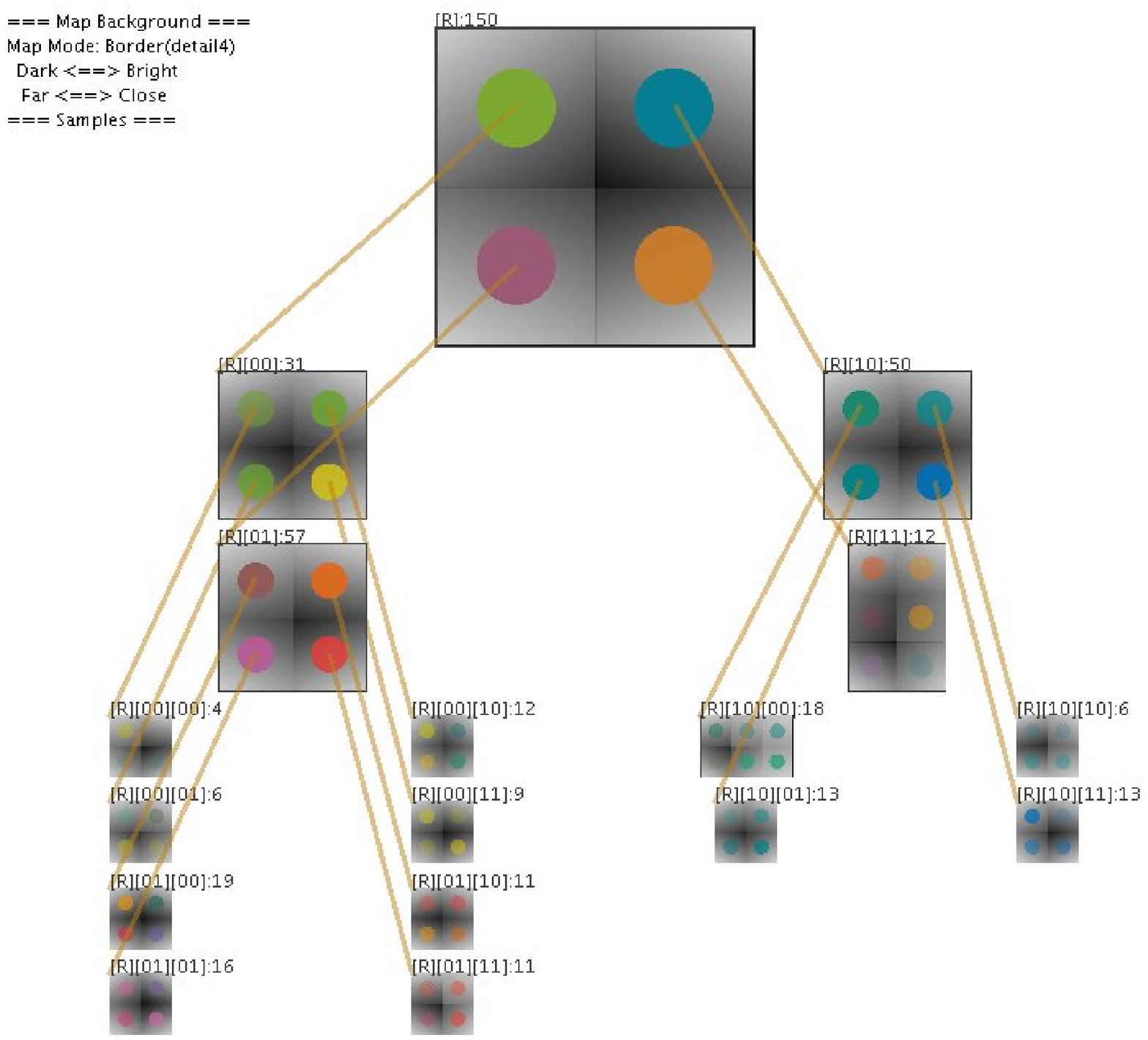}
\label{fig:GHSOM_iris0}
}
\subfigure[A Detailed Description]{
\includegraphics[scale=0.5]{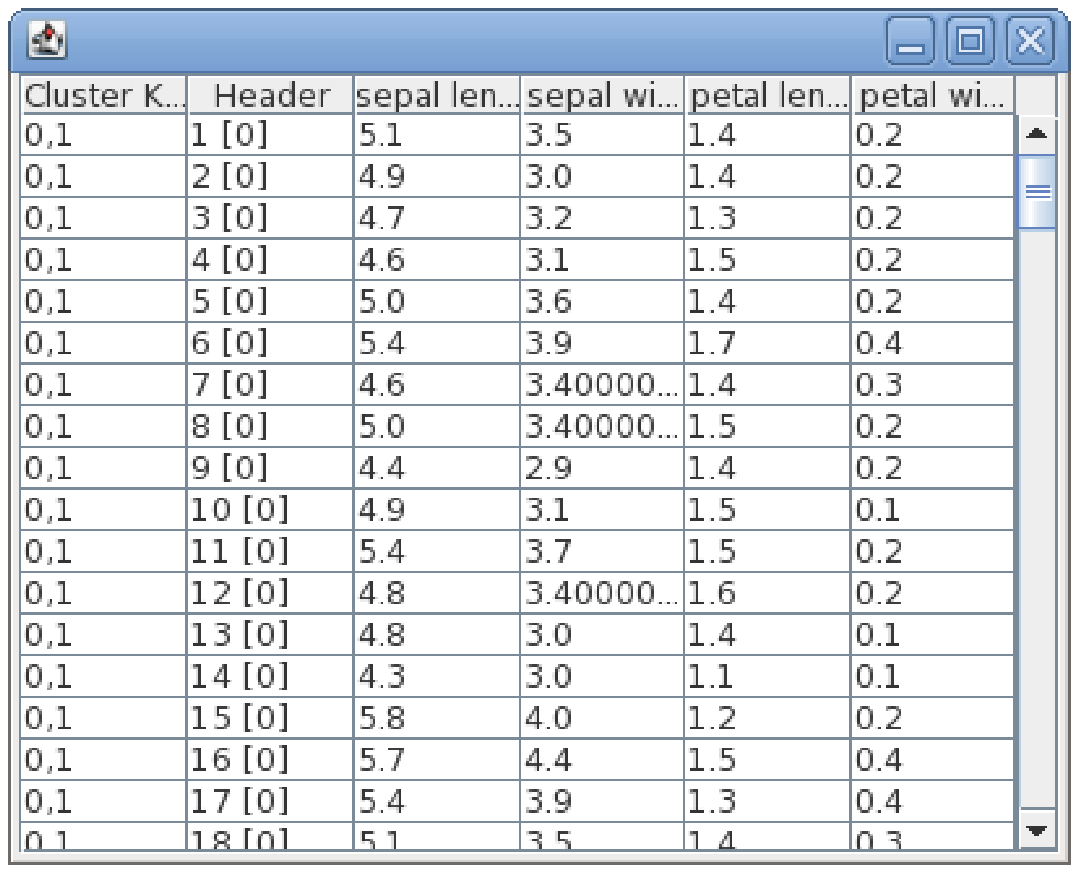}
\label{fig:GHSOM_iris_table}
}
\caption{Simulation Result for Iris data set}
\label{fig:GHSOM_iris}
\end{center}
\end{figure}

We developed the Android smartphone based interface of interactive GHSOM to acquire the knowledge intuitively. This tool was developed by Java language. Fig. \ref{fig:GHSOM_iris0} shows the clustering results of Iris dataset \cite{UCI_IRIS} by GHSOM. When we touch a unit in a map, the other window as shown in Fig. \ref{fig:GHSOM_iris_table} is displayed and the samples in allocated in the corresponding unit is listed in the table. The notation $[R][01][10]:12$ as shown in Fig. \ref{fig:GHSOM_iris0} represents the location of unit in the connection from the top level $[R]$. $[R]$ means a root node. The numerical value('12') shows the number of samples divided into the leaf map after the sequence of classification $[R][01][10]$. The numerical values in the brackets mean the position of units in the corresponded map. The first letter (eg. `0') and the second letter (eg. `1') are the position in the column and the row in the map(eg. `01'), respectively.

\begin{figure}[!tb]
\begin{center}
\includegraphics[scale=0.25]{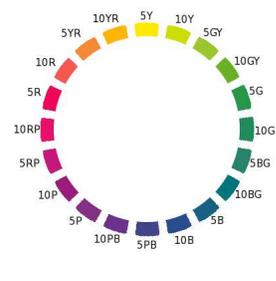}
\caption{The Munsell Color System}
\label{fig:munsell}
\end{center}
\end{figure}

The color of unit shows the pattern of sample represented in Munsell color system \cite{Munsell}, which consists of three independent dimensions: hue, value, and chroma. Munsell defined hue as ``the quality by which we distinguish one color from another.'' He selected 5 principle colors: red, yellow, green, blue, and purple; and 5 intermediate colors: yellow-red, green-yellow, blue-green, purple-blue, and red-purple. He arranged theses colors in a wheel measured off in 100 compass points as shown in Fig.\ref{fig:munsell}. Fig.\ref{fig:munsell} shows the hues of the Munsell color system at varying values, and maximum chroma to stay in the RGB gamut. A color circle is an abstract illustrative organization of color hues around a circle. The HSV color spaces are simple geometric transformation of the RGB cube into cylindrical form by using Eq.(\ref{eq:RGBtoHUE}).

\begin{equation}
\tan(h_{rgb})=\frac{\sqrt{3}(G-B)}{2R-G-B},
\label{eq:RGBtoHUE}
\end{equation}
where $h_{rgb}$ is the hue angle. The component values of R, G, and B in this paper are stored as integer numbers in the range 0 to 255, that a single 8-bit byte can provide. The component values are calculated by principal component analysis (PCA). There are some the eigen vectors and eigen values in data set. The 3 eigen vectors with the higher eigen value are selected to convert the feature of data set into the range of RGB color space. Certainly, the measure related to the quantization error is required to display colored maps effectively in the classification process. However, the size of error in the first trial is quite different from that in the reclassification. In this paper, the size of error by the PCA method is used.

 The similar color of units represents an intuitive understanding of similar pattern of samples. If the number of units in a map are increased, only a few samples could be classified into a new generated unit. Once the unit connected to the map is selected, the method re-calculates to find an optimal set of weights in the local tree structure search and then a better structure is depicted.

When the corresponding unit is touched, the system calculates the 4 allocated samples. The system continues to classify again till the user determines the GHSOM structure. We call the method the interactive GHSOM in the interactive process. The calculation result by the interactive GHSOM as shown in Fig. \ref{fig:GHSOM_iris0} is obtained and the effectiveness of the interactive GHSOM is shown as results of empirical studies.

The classification by GHSOM shows the tree structure of clusters and the connection among them. The detailed knowledge cannot be represented in the form of If-Then rules. Despite low resolution in knowledge representation, we grasp the rough sketch of knowledge structure because GHSOM shows the samples divided into each unit on the map as shown in Fig. \ref{fig:GHSOM_iris0}. Moreover, there is only a few samples in each unit. Therefore, knowledge discovery is executed by grasping the structure and a grain of knowledge.

\section{Tourist Subjective Data in MPPS}
\label{sec:experiments}

\subsection{Tourist subjective data}
\begin{figure}[tbp]
\begin{center}
\subfigure[Register a new sightseeing spot]{
\includegraphics[scale=0.15]{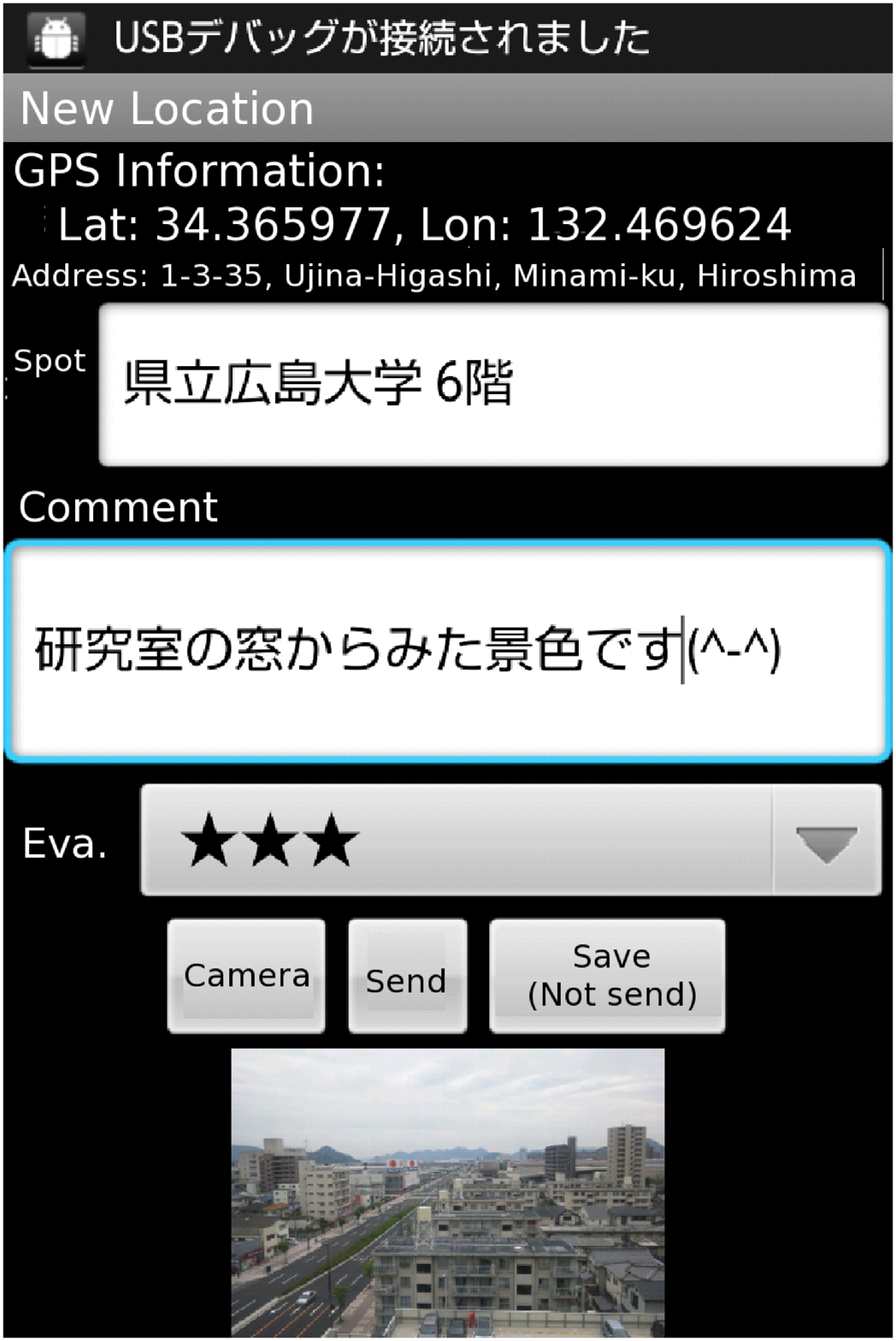}
\label{fig:Android_KankouMap_newlocation}
}
\subfigure[Google map with new locations]{
\includegraphics[scale=0.15]{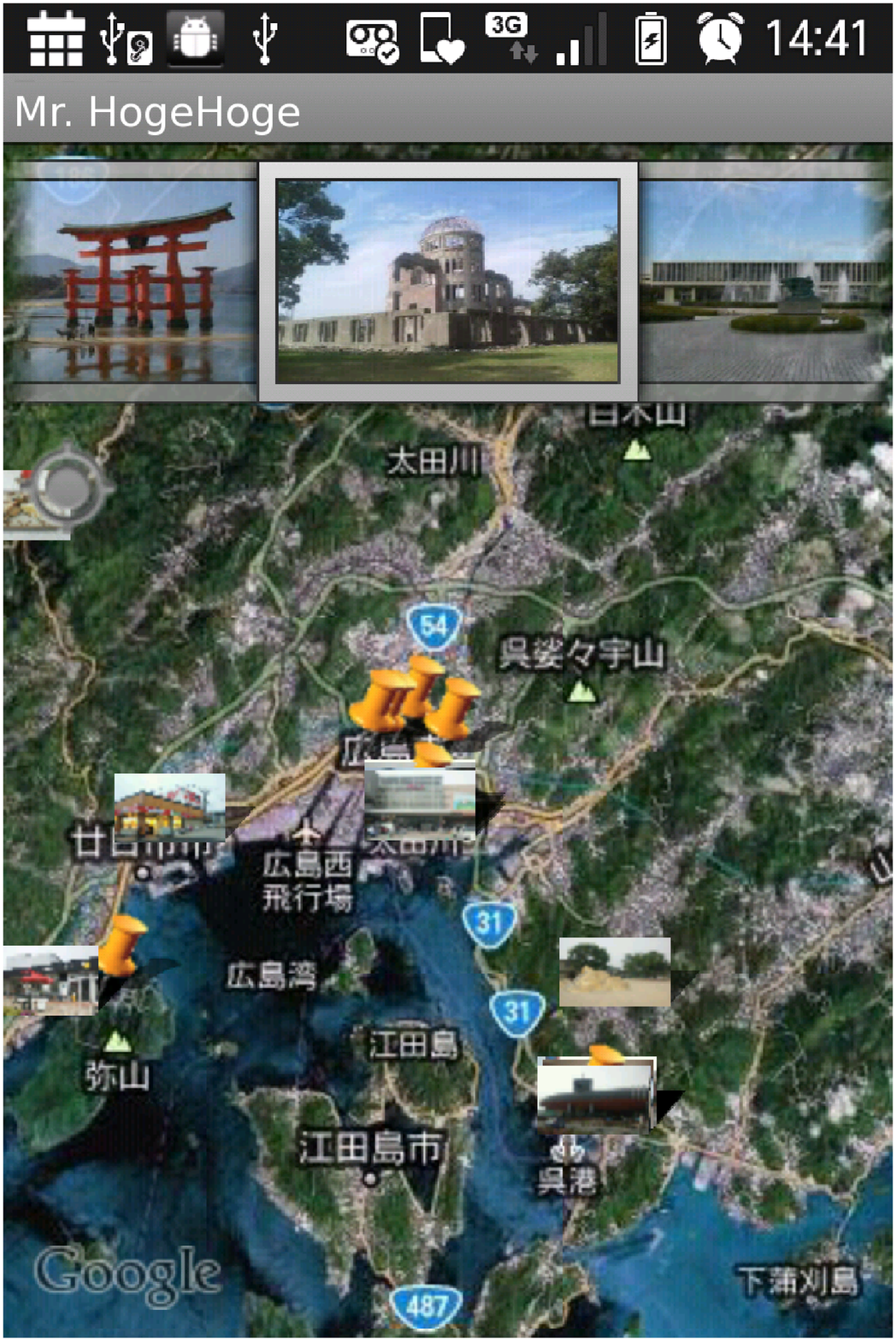}
\label{fig:Android_KankouMap_googlemap}
}
\caption{Android smartphone application `KankouMap'}
\label{fig:Android_KankouMap}
\end{center}
\end{figure}

Participation of mobile phone users in sensorial data collection both from the individual and from the surrounding environment presents a wide range of opportunities for truly pervasive applications \cite{Lane2010}. For example, the technology for determining the geographic location of cell phones and other hand-held devices is becoming increasingly available. Our developed Android smartphone application \cite{Android_Market} can collect the tourist subjective data in the research field of MPPS as shown in Fig. \ref{fig:Android_KankouMap_newlocation}. The collected subjective data consist of jpeg files with GPS, geographic location name, the evaluation of $\{0, 1, 2, 3, 4\}$ and comments written in natural language at sightseeing spots to which a user really visits. The application must obtain GPS data before taking a picture so that the pictures provide evidence to prove that the tourist visited there. Fig. \ref{fig:Android_KankouMap_googlemap} shows a Google map on which putting their pictures in the KML format. More than 500 subjective data are stored in the database through MPPS. Table \ref{tab:KankouMap} shows some samples of sightseeing spots collected by our developed MPPS.

\begin{table}[!tb]
\caption{Sample records in our developed MPPS}
\label{tab:KankouMap}
\vspace{-3mm}
\begin{tabular}{rccclc}
\hline\noalign{\smallskip}
No. & Lat. & Lon. & Alt. & Name & Eva. \\
\noalign{\smallskip}\hline\noalign{\smallskip}
6 & 34.363369 & 132.470307 & 32.30  & Oyster Street   & 2 \\
9 & 34.484011 & 132.269203 & 258.8 & Fishing Lake     & 3 \\
10 & 34.484362 & 132.269326 & 272.6 & Fishing Lake     & 4 \\
11 & 34.473791 & 132.240430 & 356.2 & Rodge            & 1 \\
13 & 34.367706 & 132.175777 & 357.5 & Futae Yaki(Cake) & 4 \\
16 & 34.388838 & 132.103882 & 575.7 & Spa Rakan        & 4 \\
58 & 34.393745 & 132.436148 &  41.4 & Game spot        & 3 \\
200 & 34.387643 & 132.430239 &  50.7 & Tomato noodle    & 4 \\
227 & 34.410682 & 133.197108 & 174.8 & Onomichi         & 4 \\
%241 & 34.393464 & 132.459653 &  52.3 & \parbox{9zw}{High quality Japanese Restaurant}& 4 \\
241 & 34.393464 & 132.459653 &  52.3 & \parbox{2.4cm}{High quality Japanese Restaurant}& 4 \\
\noalign{\smallskip}\hline\noalign{\smallskip}
\end{tabular}
\end{table}

 Because most of user comments as shown in Table \ref{tab:user_comment}, which are translated into English for easy comprehension, are short messages within 140 letters such as Twitter, the supplement information should be retrieved from web such as tourism association websites and tourism blogs as shown in Table \ref{tab:Websitelist}. The `words' in Table \ref{tab:Websitelist} means the number of words extracted from html files in the website list. The term frequency in the subjective comments as shown in Table \ref{tab:user_comment} is calculated by TF-IDF (term frequency inverse document frequency) method \cite{TF_IDF}.

 The TF-IDF calculates a weight often used in information retrieval and text mining. The term frequency $tf(t,d)$ gives a measure of the importance of the term $t$ within the particular document $d$. The inverse document frequency $idf(t)$ is a measure of the general importance of the term. A high weight in $tfidf$ is reached by a high term frequency and a low document frequency of the term in the whole collection of documents.

\begin{table}[!tb]
\caption{User comments at sightseeing spot}
\label{tab:user_comment}
\vspace{-3mm}
\begin{tabular}{rlll}
\hline\noalign{\smallskip}
%No. & User comments & \parbox{4zw}{Max of TFIDF} & \parbox{4zw}{Sum of TFIDF}\\
No. & User comments & \parbox{1cm}{Max of TFIDF} & \parbox{1cm}{Sum of TFIDF}\\
\noalign{\smallskip}\hline\noalign{\smallskip}
6 & A posh cafe is over there! & 0.133702 & 0.039891\\
%9 & \parbox{15zw}{A peaceful fishing lake. After enjoying fishing, we must eat them.} & 0.101753 & 0.026962\\
9 & \parbox{4.2cm}{A peaceful fishing lake. After enjoying fishing, we must eat them.} & 0.101753 & 0.026962\\
%10 & \parbox{15zw}{I caught some fishes. 'Yamame' is delicate.} & 0.215457 & 0.037928\\
10 & \parbox{4.2cm}{I caught some fishes. 'Yamame' is delicate.} & 0.215457 & 0.037928\\
11 & There is nothing. & 0.176589 & 0.031254\\
13 & 'Futae-Yaki' is a kind of fried cake. & 0.003162 & 0.001583\\
16 & This spa stands by roadside station. & 0.107662 & 0.025609\\
58 & Famous game center. & 0.007618 & 0.003267\\
%200 & \parbox{15zw}{Tomato Ramen is a salt ramen with tomato.} & 0.011355 & 0.004652\\
200 & \parbox{4.2cm}{Tomato Ramen is a salt ramen with tomato.} & 0.011355 & 0.004652\\
227 & The seto sea is very beautiful. & 0.150309 & 0.031332\\
241 & Very delicious, but too expensive... & 0.041700 & 0.015182\\
\noalign{\smallskip}\hline\noalign{\smallskip}
\end{tabular}
\end{table}

\begin{table}[!tb]
\caption{Tourism association website or blog list}
\label{tab:Websitelist}
\vspace{-3mm}
\begin{tabular}{llr}
\noalign{\smallskip}\hline\noalign{\smallskip}
\multicolumn{3}{c}{Tourism Association Website} \\
\noalign{\smallskip}\hline\noalign{\smallskip}
WebSite & URL & Words\\
Hiroshima & {\it http://www.kankou.pref.hiroshima.jp/} & 11,982\\
Kure & {\it http://www.urban.ne.jp/home/kurecci/} & 1,247\\
Hatsukaichi & {\it http://www.hatsu-navi.jp/} & 1,552\\
Onomichi &  {\it http://www.ononavi.com/} & 5,600\\
Fukuyama & {\it http://www.fukuyama-kanko.com/event/} & 3,449\\
Miyoshi & {\it http://www.kankou-miyoshi.jp/} & 879\\
Akitakata & {\it http://www.akitakata.jp/kankou/} & 2,044\\
\noalign{\smallskip}\hline\noalign{\smallskip}
\multicolumn{3}{c}{Tourism Blog Website} \\
\noalign{\smallskip}\hline\noalign{\smallskip}
WebSite & URL & Words\\
Personal Blog & {\it http://www.hiroshima-asobiba.net/} & 5,290\\
Outdoor Map & {\it http://hiroshima.o-map.com/} & 3,391\\
\noalign{\smallskip}\hline\noalign{\smallskip}
\end{tabular}
\end{table}

\begin{eqnarray}
\nonumber tf(t, d)&=&\frac{n(t,d)}{\sum_{k} n(k,d)}\\
\nonumber idf(t) &=& \log \frac{|D|}{|\{d \in D: t \in d\}|}\\
tfidf(t,d)&=&tf(t,d) \times idf(t),
\label{eq:tfidf}
\end{eqnarray}
where $n(t,d)$ is the occurrence count of a term $t$ in the document $j$. $|D|$ is the total number of documents in the corpus. $|\{d \in D: t \in d\}|$ is the number of documents where the term $t$ appears.

In this paper, $d$ and $t$ indicate the html file in Tourist websites and the representative word in the user comments of Android application, respectively.

However, this simulation designs to use only one TF-IDF value per a sample. A user does not know how many words representing tourism information are in one record and does not determine the `best' TF-IDF value. If a sample has two or more words with TF-IDF value, it is difficult to select the representative word per a sample. Because tourist subjective data in our developed MPPS relates to sightseeing spots, the words used in comments are also limited to the familiar location or gifts. The collected samples include valuable information that is known to few people. If there is a word with the maximum value of TF-IDF and it is a representative value in a sample, there was no appreciable difference among other words because most of TF-IDF value is small. In this paper, we consider that each comment has $N$ words at most and $N$ words are divided to 2 groups: the existing $l$ words with higher TF-IDF value and the remaining $N-l$ words with lower TF-IDF value. The TF-IDF value for $N-l$ words is 0. The experiment describes TF-IDF values in case of `$l=3$,' because the words with high TF-IDF values are not so much. Then, the value of TF-IDF field as shown in Table \ref{tab:user_comment} denotes the sum of TF-IDF value for $l$ words in a comment.

\begin{figure}[tbp]
\begin{center}
\subfigure[The classification result by the traditional GHSOM]{
\includegraphics[scale=0.4]{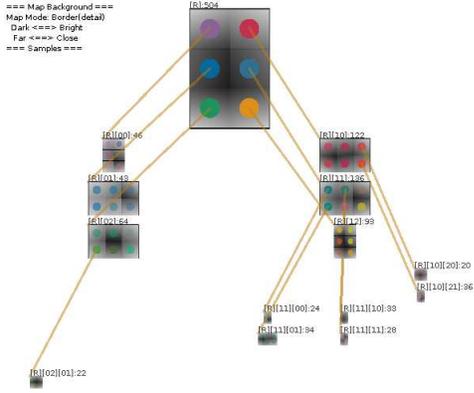}
\label{fig:GHSOM_kankoumap1}
}
\subfigure[The classification result by the interactive GHSOM]{
\includegraphics[scale=0.4]{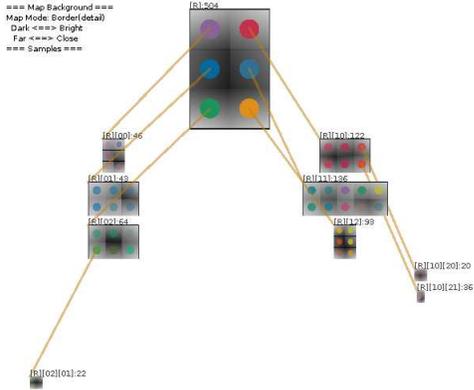}
\label{fig:GHSOM_kankoumap2}
}
\caption{Simulation result of tourist subjective data of MPPS}
\vspace{-5mm}
\label{fig:GHSOM_kankoumap}
\end{center}
\end{figure}

\subsection{Data representation in the interactive GHSOM}
Fig. \ref{fig:GHSOM_kankoumap} shows the classification results for tourist subjective data by the interactive GHSOM. The classification result in Fig. \ref{fig:GHSOM_kankoumap1} shows an example by the traditional GHSOM. 
 When the node on the map $[R][11]$ was clicked, the interactive GHSOM calculated to classify samples in the node again and the upper map connected to the node was rearranged by the insertion of new nodes as shown in Fig. \ref{fig:GHSOM_kankoumap2}. In the experiment, the parameter setting was $\tau_{1}=0.1$, $\tau_{2}=0.01$, $\alpha=0.03$, $\beta=2$.

\begin{figure}[tbp]
\begin{center}
\subfigure[Tree structure of C4.5 (before)]{
\includegraphics[scale=0.6]{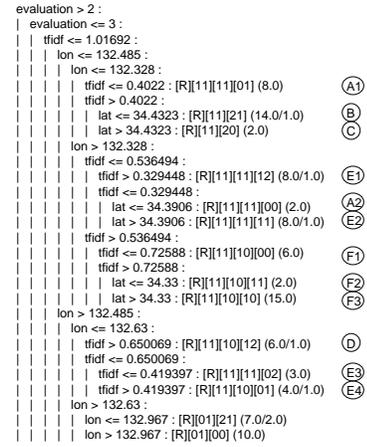}
\label{fig:C4.5_TFIDF_kankoumap1}
}
\subfigure[Tree structure of C4.5 (after)]{
\includegraphics[scale=0.6]{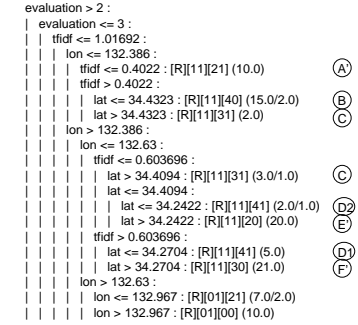}
\label{fig:C4.5_TFIDF_kankoumap2}
}
\caption{Knowledge part changed by the interactive GHSOM}
\vspace{-5mm}
\label{fig:C4.5_TFIDF_kankoumap}
\end{center}
\end{figure}

Fig. \ref{fig:C4.5_TFIDF_kankoumap} shows the tree structure of knowledge extracted by C4.5 from GPS, evaluation, TF-IDF for their comments supplemented with websites, and the clustering result by GHSOM. The prior structure as shown in Fig. \ref{fig:C4.5_TFIDF_kankoumap1} became simple knowledge as shown in Fig. \ref{fig:C4.5_TFIDF_kankoumap2} after the interactive GHSOM implementation. The symbols such as $\textcircled{\scriptsize A}$, $\textcircled{\scriptsize B}$ in Fig.\ref{fig:C4.5_TFIDF_kankoumap1} show the samples classified into the unit $[R][11]$. The knowledge is extracted in the view of longitude, user's estimated value, and TF-IDF values. However, rules in the form of multi-stage reasoning are extracted and same attribute in the antecedent part of the rule are appeared. For example, the first condition was the position of longitude, the second condition of TF-IDF values was appeared and the last condition was the criteria of longitude again. This representation is not easy comprehension for users, because more than 2 conditions of longitude are required to classify the same samples. On the other hand, there is a rule group related to the same sightseeing spot where many users submitted the tourist information. In this case, more detailed classification with the other attributes is required.

After the interactive GHSOM implementation, the knowledge parts denoted by $\textcircled{\scriptsize A1}$ and $\textcircled{\scriptsize A2}$ as shown in Fig. \ref{fig:C4.5_TFIDF_kankoumap1} was well arranged into the part in $\textcircled{\scriptsize A'}$s shown in Fig. \ref{fig:C4.5_TFIDF_kankoumap2}. On the other hand, the part in $\textcircled{\scriptsize D}$ is split into the knowledge parts denoted by $\textcircled{\scriptsize D1}$ and $\textcircled{\scriptsize D2}$ as shown in Fig. \ref{fig:C4.5_TFIDF_kankoumap2}

 These 136 samples divided into the 4 maps in the lower layer were allocated in the unit $[R][11]$ before the interactive GHSOM implementation as shown in Fig. \ref{fig:C4.5_TFIDF_kankoumap1}, because the method classified not only the corresponding node but also the samples included in the above map connected to the node simultaneously. As a result, one condition of classification knowledge arranged some granular parts of knowledge and the overall structure of knowledge can be a lucid expression. Note that the simulation result described here depends on the trials.

The complex condition such as a result is often seen in the classification by using softcomputing methodologies. Our proposed interactive GHSOM can sometimes classify the partially knowledge into the distinct subcategories and sometimes arrange the other way around, because the interactive GHSOM can classify the samples allocated into the specified node. Then, the complex structure of knowledge becomes a clear. Such results appeared in other empirical studies show that the GHSOM tool can give the intuitive graphical comprehension for real world data such as tourist subjective data in MPPS.

\subsection{Filtering Rules of Twitter}
\begin{figure}[!tb]
\begin{center}
\includegraphics[scale=0.3]{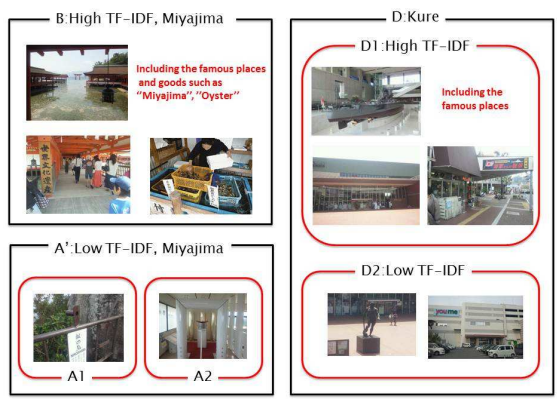}
\vspace{-5mm}
\caption{The overview of Filtering rules of Twitter}
\label{fig:Twitter_rules}
\end{center}
\end{figure}

This subsection describes the filtering rules of tourist subjective data to be tweeted. Fig. \ref{fig:Twitter_rules} shows the overview of the clustering results by the interactive GHSOM and the decision tree by C4.5 via our developed Android Application. The following filtering rule is generated according to the acquired knowledge. The idea of rule is drawn in the left side of Fig.\ref{fig:Twitter_rules}.
\begin{verbatim}
Rule:
 IF `evaluation' > 2 & `lon' < 132.386
    & `lat' < 34.4323 & `tfidf' > 0.4022
 THEN Area is `Miyajima'.
\end{verbatim}
If the message to be tweeted matches the rule, the corresponding tourist information is important and the message is outgoing on social networking service and microblogging service. The output messages with the hush tag (\#KankouMap) are appeared in the Twietter as the specified user.

\section{Conclusive Discussion}
\label{sec:conclusion}
The interactive GHSOM can restrain the growing of hierarchy in GHSOM by reforming the map in each layer interactively. The interactive GHSOM and its graphical tool can give intuitive comprehension of two or more dimensional data. In order to verify the effectiveness of our proposed method, we developed Android smartphone based Participatory Sensing system for Hiroshima tourism information which can collect the tourist subjective data in sightseeing spots. As computational results, the tree structure by GHSOM almost equals the knowledge structure extracted by C4.5. The differences between classification results as shown in Fig.\ref{fig:C4.5_TFIDF_kankoumap} are represented knowledge in If-Then rule form. The knowledge obtained from the calculation result by GHSOM is complex structure because the tree is nesting under the same condition without the numerical value of criteria. After the interactive GHSOM implementation, the obtained knowledge represents in the clear structure intuitively. The GHSOM analysis tool on the Android smartphone is expected to be solved the problems in real world and within real time, because Android is equipped with the easy human computer interaction.

Located in Hiroshima, which is remote from big cities, it is well-known widely as a sightseeing spot of World Heritage Site and a lot of guests seem to come all the way from Kyoto, Osaka, and Tokyo. Certainly, Hiroshima has two spots of UNESCO World Heritage Site. However, the recent utilization of local resources for tourism makes no difference to our desire to travel, because the means of transport has been developed from downtown to peripheral suburbs. Not only tourism association but the local citizens should give the innovative and attractive information in sightseeing to visitors. Our developed clustering system and MPPS for tourism information can discover valuable information in the unknown user subjective data and spread the information on social networking service. However, the analysis of the main subjective data such as pictures does not implemented. We will develop the image analysis method in tourism information in future.

% that's all folks

\begin{thebibliography}{1}

\bibitem{Lane2010}
Lane, N.D., Miluzzo, E., Hong L., Peebles, D., Choudhury, T., Campbell, A.T. (2010)
`A survey of mobile phone sensing', 
{\it IEEE Communications Magazine},
Vol.48, No.9, pp.140-150.

\bibitem{Android_Market}
ITProducts (2011)
`Hiroshima Tourist Map',
{\it https://market.android.com/details? id=jp.itproducts.KankouMap},
Retrieved 2011-11-15.

\bibitem{Ichimura2011a}
Ichimura, T., Yamaguchi, T. (2011)
`A Proposal of Interactive Growing Hierarchical SOM',
{\it Proc. of 2011 IEEE SMC2011},
pp.3149-3154.

\bibitem{Quinlan96}
Quinlan, J.R. (1996)
`Improved use of continuous attributes in c4.5',
{\it Journal of Artificial Intelligence Research},
No.4, pp.77-90.

\bibitem{Kohonen95}
Kohonen, T. (1995)
`Self-Organizing Maps', {\it Springer Series in Information Sciences},
Vol.~30, Springer.

\bibitem{Rauber02}
Rauber, A., Merkl, D., Dittenbach, M., 
`The growing hierarchical self-organizing map: exploratory analysis of high-dimensional data',
{\it IEEE Transactions on Neural Networks}, vol.13, pp.1331-1341.

\bibitem{UCI_IRIS}
Fisher (1936)
`Iris Dataset, UCI Machine Learning Repository',
{\it http://archive.ics.uci.edu/ml/datasets/Iris}.
Retrieved 2011-11-15.

\bibitem{Munsell}
Kuehni, R.G. (2002)
`The early development of the Munsell system',
{\it Color Research and Application},
Vol.27, No.1, pp.2027.

\bibitem{TF_IDF}
Wu, H.C., Luk, R.W.P., Wong, K.F., Kwok, K.L. (2008)
`Interpreting TF-IDF term weights as making relevance decisions',
{\it ACM Transactions on Information Systems},
Vol.26, No.3, pp.137, 2008.

\bibitem{Ichimura2012}
Ichimura, T., Kamada, S., Kato, K. (2012)
`Analysis of Tourist Subjective Text Data in Smartphone based Participatory Sensing System by Interactive Growing Hierarchical SOM'
Proc. of KES-IDT2012, Vol. 2, SIST 16, pp. 225–235, 2012.

\end{thebibliography}
\end{document}